\documentstyle[twocolumn,aps,epsfig,prl]{revtex}

\begin{document}
\bibliographystyle{prsty}
\title{Time evolution and squeezing of the field amplitude in cavity QED}
\author{J. E. Reiner, W. P. Smith, L. A. Orozco}
\address{Dept. of Physics and Astronomy,
State University of New York, Stony Brook, NY 11794-3800}
\author{H. J. Carmichael}
\address{Dept. of Physics, University of Oregon, Eugene, OR 97403-1274}
\author{P. R. Rice}
\address{Dept. of Physics, Miami University, Oxford, OH 45056}
\date{\today}
\maketitle
\begin{abstract}
We present the conditional time evolution of the electromagnetic
field produced by a cavity QED system in the strongly coupled
regime. We obtain the conditional evolution through a
wave-particle correlation function that measures the time
evolution of the field after the detection of a photon. A
connection exists between this correlation function and the
spectrum of squeezing which permits the study of squeezed states
in the time domain. We calculate the spectrum of squeezing from
the master equation for the reduced density matrix using both the
quantum regression theorem and quantum trajectories. Our
calculations not only show that spontaneous emission degrades the
squeezing signal, but they also point to the dynamical processes
that cause this degradation.
\end{abstract}
\vspace{0.1in}




\section{Introduction}
Squeezing experiments focus on the control and study of quantum
fluctuations. In a squeezed state there is a redistribution of the
fluctuations between the different quadratures of the
electromagnetic field. The modification is measurable through the
spectrum of squeezing and its comparison with that of the vacuum
state. In this paper we analyze the time evolution of the quantum
fluctuations of the electromagnetic field, the complementary
picture of the spectral density studied in squeezing. We use a new
formalism based on a third order correlation function of the
field. This wave-particle correlation combines conditional
measurements with the process of quantum measurement. The quantum
fluctuations are related to the measurement process and, in a
strongly coupled system, those fluctuations can be orders of
magnitude larger than the steady state.

By combining the wave and particle aspects of light, this
correlation function brings a new theoretical \cite{carmichael00}
and experimental \cite{foster00,foster00t} understanding to the
study of the quantum properties of squeezed light. This is because
the wave-particle correlation function and the spectrum of
squeezing form a Fourier transform pair when third order moments
of the field fluctuations are neglected. We can now analyze the
conditional time evolution of the fluctuations that give rise to
the redistribution of the noise between the two quadratures of the
electromagnetic field.

Work on the spectrum of squeezing is a well established area of
research \cite{kimble87,loudon87}. The interaction between a
single cavity mode of the electromagnetic field and several two
level atoms produce non-classical fields which exhibit squeezing.
Work on this system started with the problem of optical
bistability (OB) \cite{lugiato84,reid86,raizen87,orozco87,hope92}.
These OB systems are treated in the limit of small noise, such
that the fluctuations are only a minimal perturbation on the
steady state. More recent work has focused on a regime where there
is no small size parameter and fluctuations dominate the behavior
of the system \cite{carmichael85}. This general area is cavity QED
\cite{berman94}. While there are experimental realizations in the
microwave and optical regimes, our discussion will only consider
the latter. In both systems the reversible coupling between the
atoms and the cavity mode is larger than the irreversible loss of
coherence from spontaneous emission or cavity decay. A number of
optical studies, experimental and theoretical, have focused on the
non-classical features as observed in the intensity correlations,
{\it eg.}~photon antibunching
\cite{rempe91,mielke98,fosterpra00,clemens00}. Our recent
observations look, however, at the time dependence of the
conditional fluctuations of the field \cite{foster00} and its
connection with the spectrum of squeezing.

We found from experiment that the spectrum of squeezing was
degraded when the conditional fluctuation oscillated around a
value different from the steady state field amplitude. In order to
understand this behaviour we apply the time domain formalism to
the strong coupling regime of cavity QED. To keep the discussion
manageable we only treat the case of one or two atoms fixed in
space. The work follows the general philosophy established in
Refs. \cite{carmichael91,brecha99} where cavity QED is analyzed
without a small parameter.

The paper is organized as follows. In section II, we describe the
wave-particle correlation function and its connection to the
spectrum of squeezing. Section III discusses the properties of the
cavity QED system. Section IV shows the different techniques
utilized to calculate the spectrum of squeezing and the
wave-particle correlation function. We present our results in
section V, the discussion of those results in section VI, and our
conclusions in section VII.

\section{The wave-particle correlation function}

Figure 1 shows the apparatus that measures the correlation between
the amplitude and the intensity of the electromagnetic field. The
particle (photon) produces a trigger `click' in an avalanche photo
diode (APD) and conditionally the wave (electromagnetic field
amplitude) gets recorded in the photocurrent output of a balanced
homodyne detector (BHD). The observations of Ref.~\cite{foster00}
consider a conditioned field evolution described by a third order
correlation function of an electromagnetic field mode $\hat b$
(signal) that has a non-zero, steady state average $\langle \hat b
\rangle = \lambda$. The measured correlation is

\begin{equation}
h_\theta(\tau)=\frac{\langle:\!(\hat b^\dagger\hat b)(0)\hat
Q_\theta(\tau)\!:\rangle}{\sqrt{\eta}\lambda\langle\hat
b^\dagger\hat b\rangle}, \label{eqn:prop}
\end{equation}
where $\hat Q_\theta\equiv(\hat b {\rm{exp}}(-i\theta)+\hat
b^\dagger {\rm{exp}}(i\theta))/2$ is the quadrature amplitude
selected by the local oscillator phase $\theta$ and $\eta$ is the
coupling efficiency into the BHD. Carmichael {\it et al.} have
generalized Eq.~(\ref{eqn:prop}) to any source
\cite{carmichael00}, but for the purpose of this paper it is
sufficient to limit the discussion to the detection scheme of
Fig.~1.

We separate the fluctuations from the mean field writing,

\begin{equation}
\hat b=\lambda {\rm{exp}}(i\theta)+\Delta\hat b
\end{equation}
The correlation function can be rewritten in terms of the field
fluctuations. In the limit of Gaussian fluctuations we may neglect
third-order moments of the field fluctuations and the correlation
function is \cite{foster00}%

\begin{equation}
h_{0^{\circ}}(\tau)=1 + 2\frac{\langle:\!\Delta\hat
Q_{0^{\circ}}(0) \Delta\hat
Q_{0^{\circ}}(\tau)\!:\rangle}{\lambda^{2}+ \langle\Delta\hat
b^\dagger\Delta\hat b\rangle}+\xi(\tau), \label{eqn:cfun}
\end{equation}
where $\Delta\hat Q_{0^{\circ}}\equiv(\Delta\hat b +\Delta\hat
b^\dagger)/2$ and $\xi(\tau)$ is the residual shot noise with the
following correlation function

\begin{equation}
\overline{\xi(0)\xi(\tau)}=\frac{\Gamma}{16\eta N_s|\langle\hat
b\rangle|^2}{\rm{exp}}(-\Gamma\tau) \label{shotcorr}
\end{equation}
where $\Gamma$ is the BHD bandwidth. It is apparent from
Eq.~(\ref{shotcorr}) that once the conditioned field is normalized
for a fixed $|\langle\hat{b}\rangle|^2$ then the shot noise,
$\xi(t)$, is determined by the detection bandwidth and the number
of samples averaged. Therefore one must average over many start
``clicks" to obtain a good signal to noise ratio.

\begin{figure}
\leavevmode \centering \epsfxsize=3.1in \epsffile{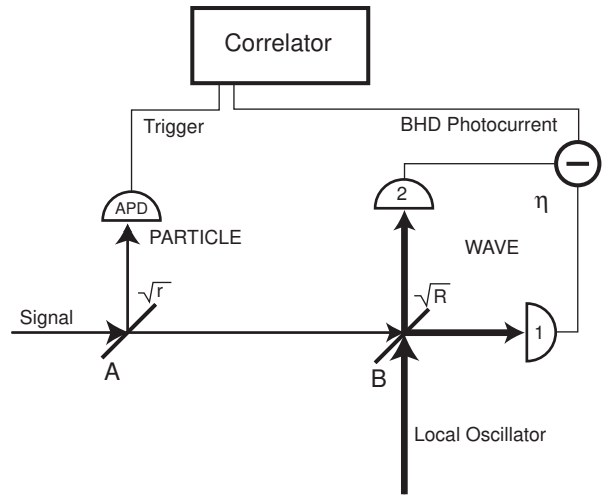}
\caption{Schematic of the wave particle correlator.} \label{fig1}
\end{figure}

The spectrum of squeezing is the cosine transform of the
normalized correlation function,

\begin{equation}
S(\theta,\nu)=4F\int_0^\infty
d\tau\cos(2\pi\nu\tau)[\overline{h_\theta(\tau)}-1],
\label{eqn:spectrum}
\end{equation}
where $F=2\kappa\langle\hat b^\dagger\hat b\rangle$ is the photon
flux into the correlator and $\overline{h_\theta(\tau)}$ is the
average of $h_\theta(\tau)$ with respect to $\xi(\tau)$.  This
average is independent of the fraction of light, $1-r$, sent to
the BHD.

The wave-particle correlator has two practical advantages over a
standard squeezing measurement. First, the fluctuations in time
reveal the entire spectrum of squeezing up to a frequency set by
the detector bandwidth. The second is that the BHD efficiency only
enters through the residual shot noise in Eq.~(\ref{eqn:cfun}).
This shot noise can be made to vanish by averaging over many
trigger starts. This implies that the spectrum of squeezing is not
degraded by imperfect detector efficiencies and propagation errors
which beset the standard squeezing measurement.

\section{Cavity QED system}

The fluctuations in the signal beam in Fig.~\ref{fig1} must be
squeezed by a non-linear source. The source we consider here
originates from a cavity QED system. The cavity QED system
consists of a single mode of the electromagnetic field interacting
with a collection of two-level atoms (see Fig.~2). Two spherical
mirrors that form a standing-wave optical cavity define the field
mode. We consider only a single or a pair of two level atoms
optimally coupled to the mode of the cavity. Dissipation plays an
important role as both the atoms and the field couple to
reservoirs. An atom can spontaneously emit light into modes other
than the preferred mode, and light inside the cavity can escape
through the mirrors. We assume that the fractional solid angle
subtended by the cavity mode is small enough so that we do not
have to make corrections to the atomic decay rates. A coherent
field, injected through one of the mirrors, drives the system. The
signal beam in Fig.~\ref{fig1} is the light that escapes from the
cavity through the output mirror.

\begin{figure}
\leavevmode \centering \epsfxsize=3.1in \epsffile{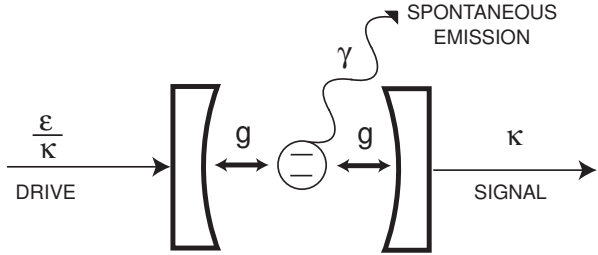}
\caption{Schematic of the cavity QED system.} \label{fig2}
\end{figure}

The Jaynes-Cummings Hamiltonian \cite{jaynes} describes the
interaction of a two-level atom with a single mode of the
quantized electromagnetic field.

\begin{equation}
\hat{H}=\hbar\omega_{\rm a} \hat{\sigma}^z +\hbar\omega_{\rm c}
\hat{a}^\dagger \hat{a} - i\hbar g (\hat{\sigma}_+ \hat{a} -
\hat{a}^\dagger \hat{\sigma}_-) \label{basicH}
\end{equation}

where $\hat{\sigma}_\pm$ and $\hat{\sigma}^z$ are the Pauli spin
operators for raising, lowering, and inversion of the atom, and
$\hat{a}^\dagger, \hat{a}$ are the raising and lowering operators
for the field. The eigenstates for Eq.~(\ref{basicH}) reveal the
entanglement between the atom and the field. The spectrum has a
first excited state doublet with states shifted by $\pm g$ from
the uncoupled resonance. The dipole coupling constant, $g$, is
given by:

\begin{equation}
g= \left( \frac {\mu^2 \omega}{2 \hbar \epsilon_0 V} \right)^{1/2}
\label{gdefined}
\end{equation}

where $\mu$ is the transition dipole moment, $\omega$ is the
transition frequency, and V is the cavity mode volume. The field
decays out of the cavity at the rate of $\kappa$ and, considering
only radiative decay, the atomic inversion decays at the rate of
$\gamma = 1/\tau$ ($\tau$ is the radiative lifetime of the atomic
transition).

From $g$, $\gamma$ and $\kappa$ we can construct two dimensionless
numbers from the OB literature that are useful for characterizing
cavity QED systems: the saturation photon number $n_{0}$ and the
single atom cooperativity $C_{1}$. They scale the influence of a
photon and the influence of an atom in the system. These two
numbers relate the reversible dipole coupling between a single
atom and the cavity mode ($g$) with the irreversible coupling to
the reservoirs through cavity ($\kappa$) and inversion decays
($\gamma$) by

\begin{equation}
C_{1}=g^{2}/\kappa \gamma
\end{equation}
and for a plane wave cavity field with the atoms sitting at the
anti-nodes,

\begin{equation}
n_{0}=\gamma^{2}/8g^{2}.
\end{equation}

They define the different regimes of OB and cavity QED. The strong
coupling regime of cavity QED, $n_{0}<1$ and $C_{1}>1$, implies
large effects from the presence of a single photon and of a single
atom in the system. Consequently, a change of one photon is a
large fluctuation in the system.

We only consider the resonant case here, with the driving field on
resonance with the atomic transition and cavity mode frequencies.
This means that we are only able to explore dynamical changes in
one quadrature. In order to generalize to other phases one must
either drive the system off resonance or add a coherent offset
like the one discussed in Ref.~\cite{carmichael00}.

Two dimensionless fields and intensities follow from the OB
literature that allow us to make contact with experiments: The
intracavity field (intensity) with atoms in the cavity $x\equiv
\langle \hat{a} \rangle/ \sqrt{n_0}$; ($X\equiv \langle
\hat{a}^\dagger \hat{a} \rangle/n_0$), and the field (intensity)
without atoms in the cavity $y \equiv {\mathcal E} / \kappa
\sqrt{n_0}$; ($Y=y^2$) where $2{\mathcal{E}}^2/\kappa$ is the
input photon flux.

\section{Calculations of the wave-particle correlation function in cavity QED}

We present the two different methods used to calculate the
wave-particle correlation function. The first is a direct
calculation from the master equation and the second is a quantum
trajectory simulation.

\subsection{Direct Calculation from the Master Equation}

A master equation of the Lindblad form can be derived for the
system shown in Fig.~\ref{fig2} using standard techniques (see for
example \cite{carmichael93}). This equation for the reduced
density operator $\hat\rho$ for the case of $N$-atoms identically
coupled to a single cavity mode reads
\begin{eqnarray}
\label{rhodef}
\lefteqn{\dot{\hat{\rho}}\equiv{\mathcal{L}}{\hat{\rho}}={\mathcal
E}[\hat{a}^\dagger -\hat{a},\hat{\rho}]+
                     g[\hat{a}^\dagger \hat{S}_- - \hat{a} \hat{S}_+,\hat{\rho}] }   \\
\nonumber & & \mbox{} + \kappa(2 \hat{a} \hat{\rho}
\hat{a}^\dagger - \hat{a}^\dagger \hat{a} \hat{\rho} - \hat{\rho}
                                 \hat{a}^\dagger \hat{a})   \\
\nonumber & & \mbox{} + \gamma/2 \sum_{j=1}^N
                 (2 \hat{\sigma}^j_- \hat{\rho} \hat{\sigma}^j_+ -
                 \hat{\sigma}^j_+  \hat{\sigma}^j_- \hat{\rho}  -
                  \hat{\rho} \hat{\sigma}^j_+ \hat{\sigma}^j_-),
\end{eqnarray}
where $\hat{S}_{\pm} = \sum_j \sigma^j_{\pm}$ are the collective
raising and lowering operators for the atoms.

Early attempts to solve this master equation are found in the OB
literature \cite{lugiato84,reid86}. They involved converting the
master equation into a $c$-number Fokker-Planck equation via a
large $N$ expansion. A connection was then made to a set of
It\^{o} stochastic differential equations. This master equation
could thus be solved analytically, but only for the case of small
fluctuations.

Our cavity QED system operates in the strongly coupled regime
where the effective number of atoms and the size of the quantum
fluctuations, relative to the steady state, will not permit us to
make any of the assumptions mentioned above. Instead we
numerically solve the master equation in a manner similar to that
described by Brecha {\it{et al}} \cite{brecha99}. We transform the
master equation into frequency space and apply the quantum
regression theorem of Lax \cite{lax63} to calculate the spectrum
of squeezing directly. We check the accuracy of our results by
increasing the maximum number of photons included in the basis,
$n_{\rm max}$, until the value of the squeezing at any given
frequency is accurate to three significant digits. This approach
permits us to calculate the spectrum of squeezing for the cavity
QED system away from the low intensity limit and it allows us to
consider the case of either one or two atoms.

\subsection{Quantum Trajectories}

Studying quantum trajectories with the wave-particle correlator
requires some justification. This section provides that
justification. We begin with a review of some of the fundamental
concepts behind quantum trajectory theory. Much of the important
work done in the field of quantum trajectories and stochastic
Schrodinger equations can be found in the review article by Plenio
and Knight \cite{Plenio} and references therein. This presentation
follows the work of Carmichael \cite{carmichael93}.

An alternative approach to studying a master equation of the
Lindblad form can be formulated with quantum trajectory theory. To
create quantum trajectories one must unravel the master equation.
We are only interested in those unravellings which correspond
directly to an experimental measurement. This measurement probes
the system by opening channels between the system and the
environment. Excitations into these channels are represented by
the following superoperator

\begin{eqnarray}
{\mathcal
S}_i\hat\rho=\hat{k}(\hat{s_i},\hat{s_i}^{\dag})\hat\rho\hat{k}^{\dag}
(\hat{s_i},\hat{s_i}^{\dag}) \label{super}
\end{eqnarray}

where ${\hat{k}}$ is a function of the creation and annihilation
operators, ${\hat{s}_i^{\dag}}$ and ${\hat{s}_i}$, for the
$i^{th}$ system operator coupled to the environment. The
probability for the system to decay through one of these channels
depends on the occupation of that particular mode,

\begin{equation}
P_i=\rm{Tr}({\mathcal S}_{\it i}\hat\rho)\it{dt}
\label{collapseprob}
\end{equation}

Quantum trajectories also provide a picture of what is happening
to the system between measurements. Equation (\ref{super})
describes how measurements are performed on the system. To see
what happens between measurements, subtract Eq.~(\ref{super}) from
Eq.~(\ref{rhodef}),

\begin{eqnarray}
({\mathcal L}-{\sum_i{\mathcal
S}_i})\hat\rho=\frac{1}{i\hbar}(\hat{H}_{\rm sys}\hat\rho -
\hat\rho\hat{H}_{\rm sys}^{\dagger}) \label{prop}
\end{eqnarray}

If we assume that the system state is initially pure then
Eqs.~(\ref{super}) and (\ref{prop}) guarantee that the conditional
density operator can be written in the factorized form
$\rho=|\psi\rangle\langle\psi|$. Therefore Eq.~(\ref{prop})
implies that a propagator exists for the system wavefunction
between measurements. This propagator is a modified, non-unitary,
Hamiltonian with a real and imaginary part, $\hat{H}_{\rm
sys}=\hat{H}_R+i\hat{H}_I$.

Below is the quantum trajectory algorithm which propagates the
system wavefunction forward in time.

1.  Choose an initial state for the system.

2. Calculate the probability for the system to decay through each
of the loss channels from Eq.~(\ref{collapseprob}).

\begin{equation}
P_i=\langle\hat{k}^{\dagger}(\hat{s}_i)\hat{k}(\hat{s}_i)\rangle
dt \label{expprob}
\end{equation}

3. Associate a uniformly distributed random number between 0 and 1
with each loss channel in step 2. If the probability for the
$i^{th}$ channel is greater than its corresponding random number
then collapse the system wavefunction,

\begin{equation}
|\psi(t+dt)\rangle=\hat{k}(\hat{s}_i)|\psi(t)\rangle
\label{eqcollapse}
\end{equation}

In the unlikely event of multiple collapses, choose only one
collapse via another random number.

4. If the probability for loss through each channel is less than
the corresponding random numbers, then propagate the system
wavefunction with the effective Hamiltonian from Eq.~(\ref{prop}),

\begin{equation}
|\psi(t+dt)\rangle=(1-\frac{\hat{H}_{\rm sys}
}{i\hbar}dt)|\psi(t)\rangle \label{eqprop}
\end{equation}

5.  Normalize the system wavefunction and repeat from step 2.

This is the basis for quantum trajectories. The master equation
describes the evolution of the reduced density matrix. A
connection is made between this density matrix and a system
wavefunction. This wavefunction leads to measurements via losses
through the system's decay channels. These measurements update our
knowledge of the system wavefunction. This allows one to study how
measurements performed on a system will, through conditioning,
affect the system state.

In order to unravel Eq.~(\ref{rhodef}) with the wave-particle
correlator we need to first add the BHD to the system. The BHD
consists of a local oscillator which is modeled by a driven,
single mode coherent field.  We write the master equation for the
system plus local oscillator mode as

\begin{eqnarray}
\label{rhodeflo} \lefteqn{\dot{\hat{\rho}}={\mathcal
E}[\hat{a}^\dagger -\hat{a},\hat{\rho}]+
                     g[\hat{a}^\dagger \hat{S}_- - \hat{a} \hat{S}_+,\hat{\rho}] }   \\
\nonumber & & \mbox{} + \kappa(2 \hat{a} \hat{\rho}
\hat{a}^\dagger - \hat{a}^\dagger \hat{a} \hat{\rho} - \hat{\rho}
                                 \hat{a}^\dagger \hat{a})   \\
\nonumber & & \mbox{} + \kappa_{LO}(2 \hat{c} \hat{\rho}
\hat{c}^\dagger - \hat{c}^\dagger \hat{c} \hat{\rho} - \hat{\rho}
\hat{c}^\dagger \hat{c}) + \kappa_{LO}\chi[\hat{c}^\dagger -
\hat{c},\hat{\rho}]   \\ \nonumber & & \mbox{} + \gamma/2
\sum_{j=1}^N
                 (2 \hat{\sigma}^j_- \hat{\rho} \hat{\sigma}^j_+ -
                 \hat{\sigma}^j_+  \hat{\sigma}^j_- \hat{\rho}  -
                  \hat{\rho} \hat{\sigma}^j_+ \hat{\sigma}^j_-),
\end{eqnarray}
where $\hat{c}^\dagger$ and $\hat{c}$ represent the raising and
lowering operators for the local oscillator mode. The coherent
field occupies a mode whose strength we denote by $\chi$. Under a
BHD scheme the beam splitter combines the signal and local
oscillator fields equally $(R=1/2)$, as in Fig. 1, to give the
following total measured fields at photodetectors 1 and 2

\begin{equation}
\hat{\mathcal{E}}_{\rm BHD^{1,2}}=\pm i\sqrt{\kappa_{\rm
LO}}\hat{c}+\sqrt{\kappa(1-r)}\hat{a} \label{hom1}
\end{equation}
while the measured field at the photon counting detector gives

\begin{equation}
\hat{\mathcal{E}}_{\rm count}=\sqrt{2\kappa r}\hat{a}.
\label{count1}
\end{equation}

We assume that the signal mode does not affect the local
oscillator mode. This implies that the density matrix separates
into a piece which corresponds to the cavity QED system and a
piece which corresponds to the local oscillator,

\begin{equation}
\hat\rho=\hat\rho_{s}|\chi\rangle\langle\chi|. \label{separate}
\end{equation}

We define the local oscillator flux in terms of the strength of
the local oscillator mode and its cavity decay rate.

\begin{equation}
f=\kappa_{\rm LO}|\chi|^2. \label{fdef}
\end{equation}

From Eq.~(\ref{super}) we construct the master equation terms
which correspond to Eqs.~(\ref{hom1}-\ref{count1}) and spontaneous
emissions.

\begin{eqnarray}
{\mathcal S}_{\rm BHD^{1,2}}\hat\rho_{s} &=&
(\pm\sqrt{f}e^{i\theta}+\sqrt{2\kappa(1-r)}\hat{a}) \nonumber \\&
&
\times\hat{\rho}_{s}(\pm\sqrt{f}e^{-i\theta}+\sqrt{2\kappa(1-r)}\hat{a}^{\dagger})
\label{homm}
\end{eqnarray}

\begin{equation}
{\mathcal S}_{\rm count}\hat{\rho}_{s} = 2\kappa
r\hat{a}\hat{\rho}_{s}\hat{a}^{\dagger} \label{countm}
\end{equation}

\begin{equation}
{\mathcal S}_{\rm spont}\hat{\rho}_{s} =
\gamma\hat\sigma_{-}\hat\rho_{s}\hat\sigma_{+} \label{spont}
\end{equation}

After subtracting Eqs.~(\ref{homm}-\ref{spont}) from
Eq.~(\ref{rhodeflo}), we arrive at an expression which corresponds
to Eq.~(\ref{prop}).

\begin{eqnarray}
({\mathcal{L}}&-&{\mathcal{S}}_{\rm BHD^{1,2}}-{\mathcal{S}}_{\rm
count}-{\mathcal{S}_{\rm spont}})\hat{\rho}_{s} = {\mathcal
E}[\hat{a}^\dagger -\hat{a},\hat{\rho}_{s}] \nonumber \\ & & +
g[\hat{a}^\dagger \hat{S}_- - \hat{a} \hat{S}_+,\hat{\rho}_{s}] -
\kappa(\hat{a}^\dagger \hat{a} \hat{\rho}_{s} +
\hat{\rho}_{s}\hat{a}^\dagger \hat{a}) \nonumber
\\ & & -\gamma/2 \sum_{j=1}^N
(\hat{\sigma}^j_+ \hat{\sigma}^j_- \hat{\rho}_{s} + \hat{\rho}_{s}
\hat{\sigma}^j_+ \hat{\sigma}^j_-) - f\hat\rho_{s} \label{propm}
\end{eqnarray}

Equation (\ref{propm}) provides a modified Hamiltonian which
propagates the system's wavefunction between measurements.
Equations (\ref{homm}) and (\ref{countm}) govern how ``clicks" at
the BHD and APD disrupt that evolution.

There are two types of detections, not including spontaneous
emissions, which occur under the conditional BHD measurement. The
first comes from Eq.~(\ref{countm}) which corresponds to
detections at the APD. According to Eq.~(\ref{collapseprob}) these
collapses occur with a probability of $2\kappa
r\langle\hat{a}^{\dagger}\hat{a}\rangle dt$ and they produce a
signifigant change in the wavefunction as is evident from
Eq.~(\ref{eqcollapse}).

The second type of collapse corresponds to the ``clicks" at the
BHD. The strength of the local oscillator tells us that there are
many BHD ``clicks" within a time interval set by $\kappa^{-1}$.
Eq.~(\ref{eqcollapse}) shows that the effect of each of these
``clicks" on the system wave function is almost negligible. This
implies that a straight forward trajectory simulation with the
algorithm described above in Eqs.~(\ref{expprob}-\ref{eqprop})
will not provide an efficient method for modeling the
wave-particle correlator. Instead we may associate a photocurrent
with the detections registered by the BHD. The superposition of
the cavity field with the local oscillator field in
Eq.~(\ref{hom1}) implies that the BHD photocurrent and the system
wavefunction evolution are not independent. With the principles
outlined in Sects. 8.4, 9.2, and 9.4 of \cite{carmichael93} we
combine Eqs.~(\ref{homm}-\ref{propm}) to simultaneously calculate
this photocurrent from the BHD setup and then evolve a
corresponding stochastic Schr\"{o}dinger equation forward in time.
This analysis leads to the following expressions for the
difference in photocurrents between photodetectors 1 and 2 in
Fig.~1, and wavefunction propagation between photodetections at
the APD and spontaneous emissions out the sides of the cavity.

\begin{equation}
di=-\Gamma(idt -
\sqrt{8\kappa(1-r)}\langle\hat{a}_\theta\rangle_cdt + dW_t)
\label{current}
\end{equation}

\begin{eqnarray}
d|\bar{\psi}\rangle_c = [\frac{H_{\rm
sys}}{i\hbar}&d&t+\sqrt{2\kappa(1-r)}\hat{a}{\rm{exp}}(-i\theta)\nonumber
\\& & \times(\sqrt{8\kappa(1-r)}\langle\hat{a}_\theta\rangle_cdt +
dW_t)|\bar{\psi}\rangle_c \label{stochschr}
\end{eqnarray}
where $\Gamma$ is the BHD bandwidth and $dW_t$ is the same Wiener
noise increment in both equations.

This BHD difference current, $i(t)$, along with a set of start
times $\{t_j\}$, form a stochastic measurement record. The source
quasimode $\hat a$ is in a quantum state $|\psi_{\rm
REC}(t)\rangle$ conditioned on the past record. We simulate
Eqs.~(\ref{current}-\ref{stochschr}) on a computer. By sampling an
ongoing realization of $i(t)$ for many ``start'' times (APD
detections realized concurrently) we may calculate the following
averaged photocurrent,

\begin{equation}
{\mathcal{H}(\tau)}=\frac1{N_s}\sum_{j=1}^{N_s}i(t_j+\tau).
\label{eqn:signal}
\end{equation}

A connection exists between the averaged photocurrent and the
wave-particle correlation function \cite{carmichael00}. In the
limit of large bandwidth this connection reads

\begin{equation}
h_{\theta}(\tau)=\frac{\mathcal{H}(\tau)}{\langle\hat{a}\rangle\sqrt{8\kappa(1-r)}}.
\label{eqn:conversion}
\end{equation}

The dual nature of the measurement process provides one of the
many strengths of the conditional field measurement. It allows the
use of quantum trajectory theory to unravel the master equation in
two distinct ways. A numerical simulation of
Eqs.~(\ref{current}-\ref{stochschr}) reproduces the experimental
results for the averaged photocurrent. We can understand some of
the mechanisms which lead to this averaged result by simulating
the algorithm described in Eqs.~(\ref{expprob}-\ref{eqprop}) and
replacing Eq.~(\ref{eqprop}) with Eq.~(\ref{stochschr}). Setting
$r\approx1$ reduces the conditional homodyne simulation to a
cleaner photocounting simulation. This leads to an understanding
of the dynamical processes which create the spectrum of squeezing.

\section {results}

The simplest parameter to change during the course of the
experiment is the strength of the driving field. This motivates
our study of the effect of driving intensity on the spectrum of
squeezing. The strength of the driving field is parameterized in
terms of the saturation photon number, $n_{\rm{o}}$. Therefore
weak fields imply $X\ll1$, for intermediate fields $X\approx1$,
and strong fields correspond to $X\gg1$. We begin by reporting the
results of a low intensity calculation with both the quantum
trajectories and the direct calculation from the master equation.
We then increase the intensity of the driving field and also the
number of atoms in the cavity from one to two. This degrades the
spectrum of squeezing in a way which we hope to understand through
the study of quantum trajectories.

\subsection{Low Intensity Results for $h(0^\circ,\tau)$ and $S(0^\circ,\nu)$}

\begin{figure}
\leavevmode \centering \epsfxsize=3.1in \epsffile{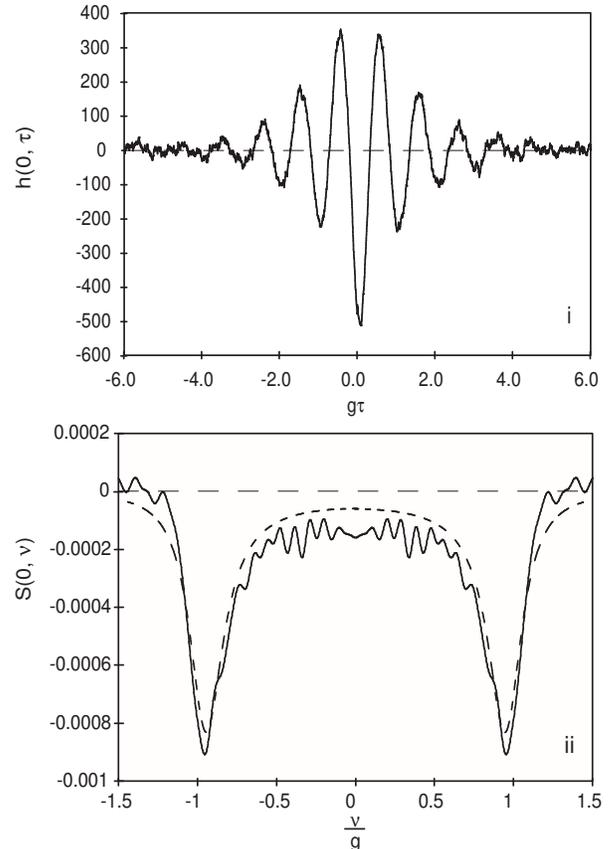}
\caption{i. Wave-particle correlation $h(0^{\circ},\tau)$ for very
low intensity excitation calculated from the quantum trajectory
implementation of the conditioned homodyne photocurrent. The
following parameters were used:
$(g,\kappa,\gamma,\Gamma)$/$(2/\pi)$ = $(38.0, 8.7, 3.0, 100)$
MHz, $X=2.99\times10^{-4}$, $r=0.5$ and $N_{\rm starts}$ =
$55000$. ii. Spectrum of squeezing calculated from the cosine
Fourier transform of $h(0^{\circ},\tau)$. The dashed line is the
spectrum of squeezing calculated directly from the master equation
with the quantum regression theorem.} \label{fig5}
\end{figure}

Figure \ref{fig5} shows the equivalence of the third order
correlation function $h(0^{\circ},\tau)$ (i) and the spectrum of
squeezing (ii) for a strongly coupled cavity QED system. The
dashed line in Fig.~\ref{fig5}ii is the spectrum of squeezing
calculated directly from the quantum regression theorem. The solid
line is the Fourier transform (see Eq.~(\ref{eqn:spectrum})) of
Fig.~\ref{fig5}i which comes from averaging the photocurrent from
a quantum trajectory simulation over 55000 ``starts" . Both
approaches show the damped Rabi oscillations which follow and
precede a photodetction. In the weak field excitation limit, Rice
and Carmichael \cite{rice88} derived an analytical expression for
the spectrum of squeezing which agrees with these results.

Detecting a photon in the APD collapses the state of the cavity
field. The oscillations in the field after a detection show the
collapsed state oscillating back towards the steady state. In the
Gaussian noise approximation the time symmetry of the measured
correlation function, Eq.~(\ref{eqn:cfun}), guarantees the
oscillation before the detection.

\subsection{From one atom to two atoms at intermediate intensity}

It has not been possible to obtain analytical expressions for the
spectrum of squeezing in the case of intermediate driving fields.
In this regime one cannot linearize the system fluctuations about
the steady state field as was done in the strong field case
\cite{reid85}. It is also impossible to perform a small parameter
expansion in the driving field as was done in the weak field case
\cite{carmichael91}. We are instead forced to rely on numerical
calculations to give us insight into a system's spectrum of
squeezing. This approach has its limitations.

\begin{figure}
\leavevmode \centering \epsfxsize=3.1in \epsffile{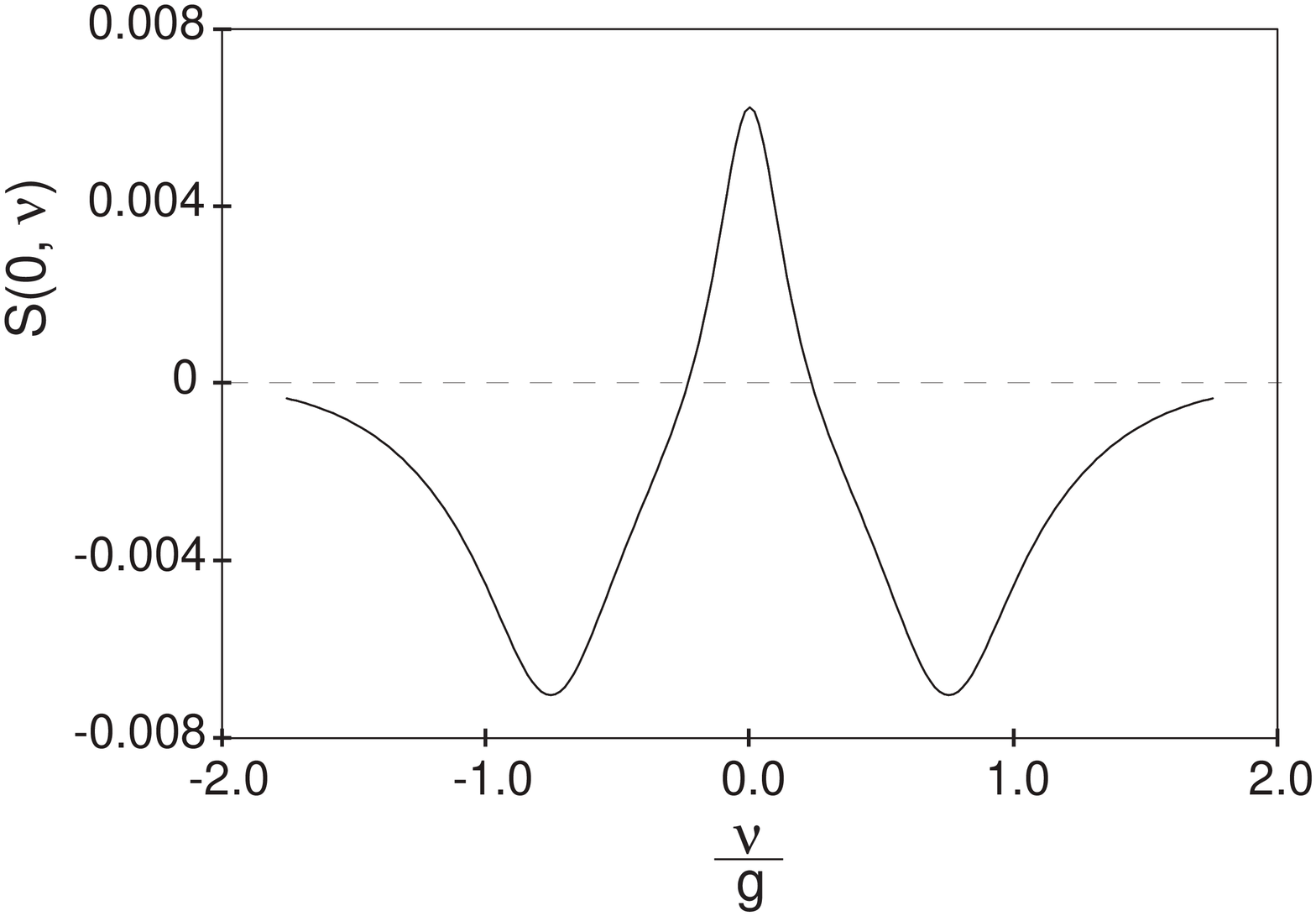}
\caption{The spectrum of squeezing calculated directly from the
master equation with the quantum regression theorem for N=1 and
high intensity. $X = 104.0$, $(g,\kappa,\gamma)$/$(2\pi)$ =
$(38.0,8.7,3.0)$ MHz.} \label{fig8}
\end{figure}

\begin{figure}
\leavevmode \centering \epsfxsize=3.1in \epsffile{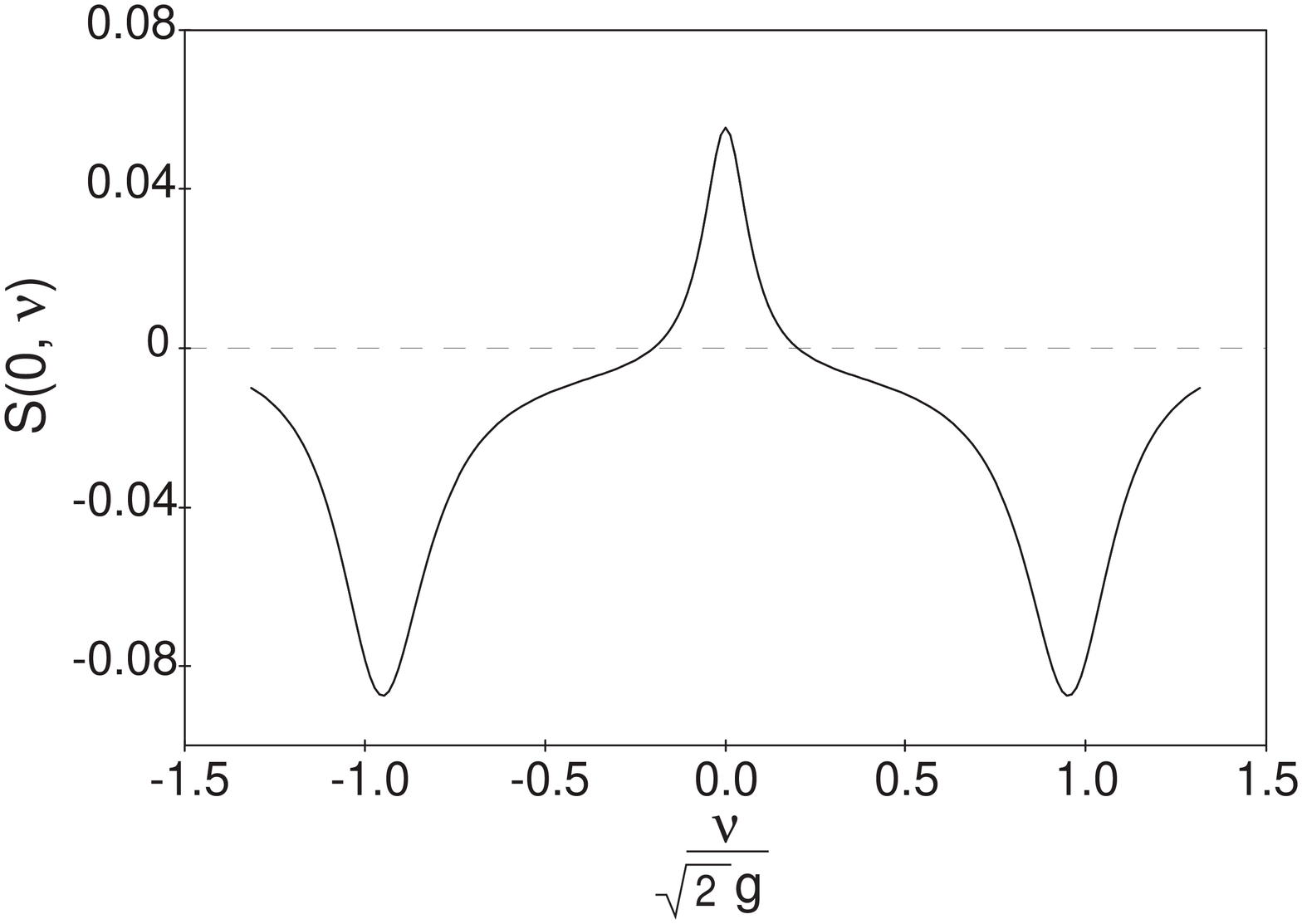}
\caption{The spectrum of squeezing calculated directly from the
master equation with the quantum regression theorem for N=2 and
intermediate intensity. $X = 1.36$, $(g,\kappa,\gamma)/(2\pi)$ =
$(38.0/\sqrt{2},8.7,3.0)$ MHz.} \label{fig7}
\end{figure}

The spectra of squeezing for increased excitation in the case of
one and two atoms have distinct quantitative differences.
Fig.~\ref{fig8} shows one such spectrum calculated for a driving
field of ${\mathcal{E}}$/$\kappa=1.50$ and one atom.
Fig.~\ref{fig7} shows the result for two atoms with
${\mathcal{E}}$/$\kappa=0.975$. We calculated both plots with the
same decay parameters but different atom field couplings, $g$ to
keep the Rabi frequency, $g\sqrt{N}$, the same.

\begin{figure}
\leavevmode \centering \epsfxsize=3.1in \epsffile{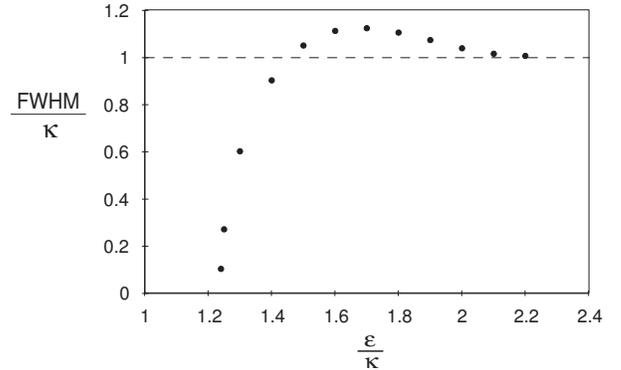}
\caption{Dependence of the width of the peak at zero frequency in
the squeezing spectrum as a function of the input driving strength
for N=1. The following parameters were used:
$(g,\kappa,\gamma)/2\pi$ = $(38.0, 8.7, 3.0)$ MHz. The FWHM has
been normalized by the cavity decay rate, $\kappa$.}
\label{fig12}
\end{figure}

The width and shape of the central peaks are very different. For
the case of the single atom it is clear, from Fig.~\ref{fig8} and
\ref{fig12}, that the cavity decay rate dominates the width. In
order to see a peak at zero frequency one must drive the one atom
system much harder than the two atom system, note that the
intracavity intensity, normalized by the saturation photon number,
is nearly $100$ times greater for the one atom case. This can be
understood from the point of view of an atom saturated with a
strong driving field. At this point most of the photons that enter
the cavity will bounce between the two mirrors without ever
interacting with this single atom. This leads to less squeezing
and a peak in the zero quadrature spectra at zero frequency with a
width which is dominated by the cavity lifetime.

\begin{figure}
\leavevmode \centering \epsfxsize=3.1in \epsffile{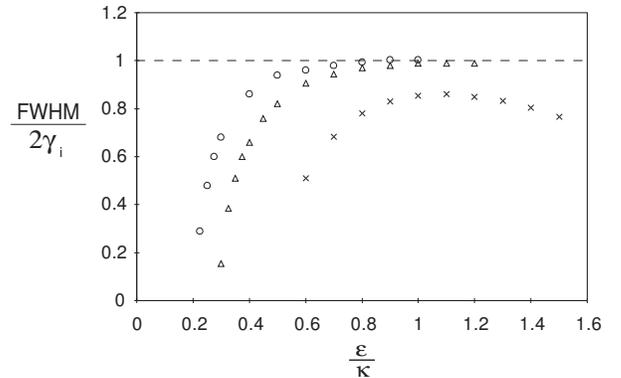}
\caption{Dependence of the width of the peak at zero frequency in
the squeezing spectrum for three different spontaneous emission
rates as a function of the input driving strength for N=2 and the
following parameters:
$(g,\kappa,\gamma_{1},\gamma_{2},\gamma_{3})/2\pi$ =
$(38/\sqrt{2},8.7,3.0,1.0,0.5)$ MHz. The FWHM for each plot has
been normalized by its respective spontaneous emission rate,
$\gamma_i$. The circles are for $\gamma=0.5$, the triangles are
for $\gamma=1.0$, and the x's are for $\gamma=3.0$. Notice the
difference in driving strengths, $\epsilon/\kappa$, between this
figure and the previous one.}
\label{fig13}
\end{figure}

When we plot the full-width at half-max (FWHM) for different
driving fields in the two atom system, Fig.~\ref{fig13}, we can
see that spontaneous emission plays a more influential role. This
is not surprising because the coupling to the environment leads to
a degradation of the squeezing signal, and these incoherent
processes appear at zero frequency in the interaction picture. The
question we hope to answer with quantum trajectories is the
following: What is the mechanism by which spontaneous emission
gives rise to this degradation in the squeezing signal?

\subsection{High Intensity}

\begin{figure}
\leavevmode \centering \epsfxsize=3.1in \epsffile{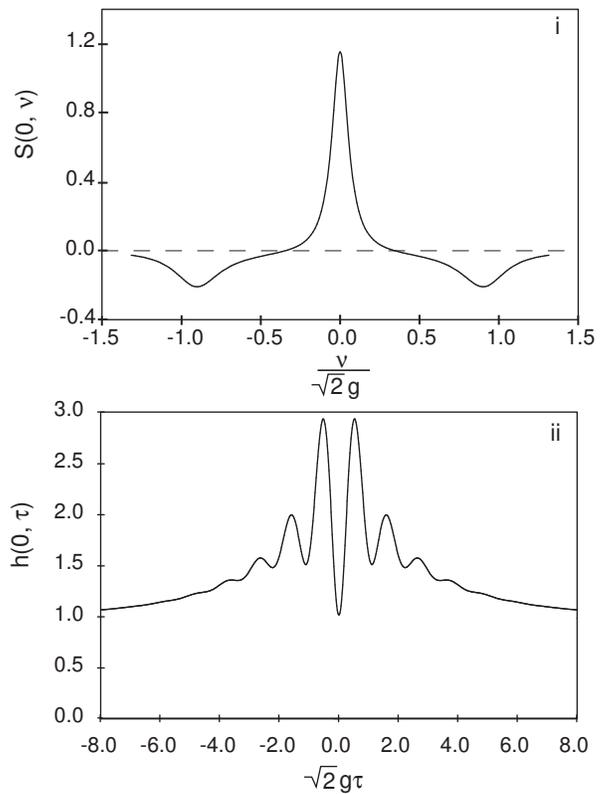}
\caption{i. Spectrum of Squeezing and ii. its inverse fourier
transform for two atom system under a stronger driving field. This
figure shows the influence of the peak at zero frequency on the
fluctuations in the time domain. $X = 18.1$ and
$(g,\kappa,\gamma)/(2\pi) = (38.0/\sqrt{2},8.7,3.0)$
MHz.}\label{fig9}
\end{figure}

A large positive peak in the spectrum of squeezing appears at zero
frequency when we increase the strength of the driving field to
the case when there is about one-tenth of a photon in the cavity,
and two atoms interacting with the mode. Negative peaks appear at
a frequency less than that of the coherent coupling, $g$, between
the atoms and the field. Figure \ref{fig9}i shows the result of
directly solving the master equation with the quantum regression
theorem.

The connection between the spectrum of squeezing and the time
domain fluctuations shows that the field undergoes an exponential
damping which is not present for the weaker driving fields (see
Fig.~\ref{fig9}ii). The careful reader may think there is a slight
problem in saying that the connection between the spectrum of
squeezing and the third order correlation function still holds in
these stronger fields. We will show, from the results of a quantum
trajectory simulation in section VI.b, that the qualitative
behaviour of the fluctuations remain the same. Therefore the third
order correlation function provides a legitimate method for
studying the spectrum of squeezing beyond the weak field
approximation.

We can get a preliminary understanding of the strong field
spectrum of squeezing from the OB work of Reid and Walls
\cite{reid85}. They showed that in the optically bistable regime
of a strongly driven, on resonance cavity, the spectrum of
squeezing is a Lorentzian centered at zero frequency. We
understand that this peak comes from the coupling of the cavity
mode to its environment, but until the recent experiments of
Foster {\it{et al.}} \cite{foster00}, we have had no motivation to
understand the origin of this peak in the time domain.

\section{discussion}

We have analyzed some of the many trajectories which go into
making up the third order correlation functions shown in the last
section by setting $r\approx1$ in Eq.~(\ref{stochschr}). We first
begin by reviewing some results in the weak field regime and then
we consider the more interesting regime of stronger driving
intensities.

\subsection{Weak Field Trajectories}

\begin{figure}
\leavevmode \centering \epsfxsize=3.1in \epsffile{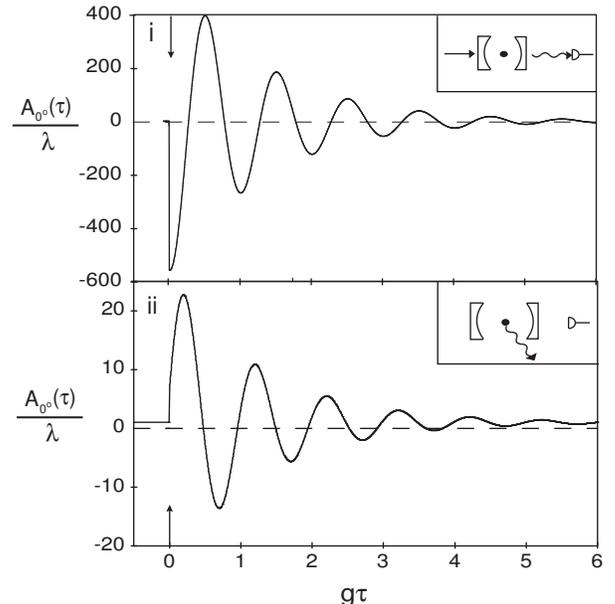}
\caption{Regression of a cavity QED system back to steady state;
N=1, low intensity i. after the detection of a photon escaping out
of the cavity mode ii. after the escape of a photon through
spontaneous emission. The inset shows the sequence of events in
terms of the cavity QED system and the detector. The parameters
used are the same as in Fig.~\ref{fig5}.} \label{fig3}
\end{figure}

Figure \ref{fig3} shows examples of the conditional field
evolution calculated quantum trajectory simulations. Figure
\ref{fig3}i shows the evolution of the field following the
detection of a photon escaping through the cavity mode and
Fig.~\ref{fig3}ii shows the field evolution following the
spontaneous emission of a photon out the side of the cavity. We
used the following realistic parameters from the work of Foster
{\it{et al}} \cite{foster00,foster00t}:
$(g,\kappa,\gamma)/2\pi$=$(38.0,8.7,3.0)$ MHz. This means that the
system is in the strong coupling regime for one and two atoms.

The basic dynamical mechanism which describes these two results
can be traced back to the change of the state $|\psi\rangle$
following a collapse operation of the type found in
Eq.~\ref{eqcollapse}. We now consider the reduction of the
equilibrium state of the cavity QED system on the occasion of a
triggering photon detection (see Fig.~\ref{fig1}). Defining $\hat
A_\theta\equiv(\hat a {\rm{exp}}(-i\theta)+\hat a^\dagger
{\rm{exp}}(i\theta))/2$, where $\hat a$ is the annihilation
operator for the cavity field and $\theta$ is the BHD phase, we
consider the quadrature amplitude, $\hat A_{0^\circ}$, in phase
with the steady state of the field $\lambda\equiv\langle\hat a
\rangle$. For weak excitation, and assuming fixed atomic positions
$\{\vec r_j\}$, to second order in $\lambda$ the equilibrium state
is the pure state \cite{carmichael91,brecha99}

\begin{eqnarray}
|\psi_{\rm{SS}}
\rangle=&&[|0\rangle+\lambda|1\rangle+(\lambda^2/\sqrt2)\alpha\beta
|2\rangle+ \cdots ]|G\rangle \nonumber \\
+&&[\phi|0\rangle+\lambda\phi\beta|1\rangle+\cdots
]|E\rangle+\cdots \label{eqn:eqstateqed}
\end{eqnarray}
where $|G\rangle$ is the one or two-atom ground state and
$|E\rangle$ is for one atom in the excited state with all others
in the ground state. We assume that all the atoms are coupled to
the cavity mode with the same strength, $g$. The explicit form for
$\phi, \alpha$ and $\beta$ follow from solving the master equation
in the steady state \cite{carmichael91}. They are

\begin{equation}
\phi=-\frac{2\sqrt{N}g}{\gamma}\lambda
\end{equation}

\begin{equation}
\alpha=1-2C_1',\label{alpha}
\end{equation}

\begin{equation}
\beta=\frac{1+2C}{1+2C-2C_1'}\label{beta}
\end{equation}
where $C\equiv NC_1$ and
$C_1'\equiv\frac{C_1}{(1+\gamma/2\kappa)}$.

After detecting the escaping photon, the conditional state is
initially the reduced state $\hat a|\psi\rangle/\lambda$, which
then relaxes back to equilibrium. The reduction and regression is
traced by \cite{carmichael91,brecha99}

\begin{equation}
|\psi\rangle\to\{|0\rangle+\lambda[1+\zeta
f(\tau)]|1\rangle+\cdots\} |G\rangle+\cdots, \label{eqn:reduced}
\end{equation}
where

\begin{equation}
\zeta=-\frac{4C_1'C}{1+2C-2C_1'}
\end{equation}

\begin{equation}
f(\tau)={\rm{exp}}(-(\kappa+\gamma/2)\tau/2)\!\left\{\cos\Omega\tau-\Phi\sin\Omega\tau\right\},
\end{equation}

\begin{equation}
\Omega=\sqrt{{Ng^2-{1\over4}\displaystyle}(\kappa-\gamma/2)^2}.
\end{equation}

\begin{equation}
\Phi=-\frac{2\kappa+\gamma}{4\Omega}
\end{equation}

Thus, the quadrature amplitude expectation makes the transient
excursion $\langle\hat A_{0^\circ}\!\rangle\to\lambda[1+\zeta
f(\tau)]$ away from its equilibrium value $\langle\hat
A_{0^\circ}\!\rangle=\lambda$.

In contrast, an undetected spontaneous emission produces the
reduced state $\hat\sigma^{-}|\psi\rangle/\lambda$, which sets up
a completely different evolution as shown in Fig.~\ref{fig3}ii.
The reduction and regression in this case is again traced by
Eq.~(\ref{eqn:reduced}) with the following replacements:

\begin{equation}
\zeta=\frac{2C_1'}{1+2C-2C_1'}
\end{equation}

\begin{equation}
\Phi=\frac{2\kappa-\gamma}{4\Omega}+\frac{2Ng^2(\frac{q\beta}{\sqrt{2}}-1)}{\gamma\Omega(\beta-1)}
\end{equation}
with $q=\sqrt{1-\frac{1}{N}}$.

These two distinct behaviors correspond fairly loosely to the
regression to equilibrium observed in the step excitation in the
field, Fig.~\ref{fig3}i, and a step excitation in the atomic
polarization, Fig.~\ref{fig3}ii. Note the phase shift between the
two responses. The steady state wavefunction determines the size
of the steps.

\begin{equation}
\frac{\langle\hat A_{0^\circ(\rm cavity)}\rangle}{\langle\hat
A_{0^\circ(\rm steady state)}\rangle}=\alpha\beta
\end{equation}

\begin{equation}
\frac{\langle\hat A_{0^\circ(\rm spontaneous)}\rangle}{\langle\hat
A_{0^\circ(\rm steady state)}\rangle}=\beta
\end{equation}

We see from Eq.~(\ref{beta}) that $\beta$ is a number greater than
zero. The field will always remain positive when an atom
spontaneously emits. As long as
$g>\sqrt{\gamma(\kappa+\frac{\gamma}{2})/2}$, then through
$\alpha$ (Eq.~\ref{alpha}), the field will change sign when a
photon escapes through the cavity mode.

The parameters used to create the plots in Fig.~\ref{fig3} are the
same as those used in Fig.~\ref{fig5}. Except for the residual
shot noise in Fig.~\ref{fig5}i, these two plots are in agreement
with each other for $\tau>0$.

\subsection{High Field Trajectories}
The weak field calculations of section VI.a make it clear that in
the strong coupling regime a cavity emission will always produce a
negative shift in the field. In order for us to observe fields of
the type in Fig.~\ref{fig9}ii and spectrum of squeezing of the
type in Figs.~\ref{fig8}, \ref{fig7} and \ref{fig9}i we must find
a mechanism which causes the expectation of the cavity field to
jump positive after a cavity emission.

It follows from Eq.~(\ref{eqn:eqstateqed}) that the ratio of the
probability for a spontaneous emission to the probability for a
cavity emission from steady state is

\begin{equation}
\frac{P_{\rm spont}}{P_{\rm cavity}}=2NC_1. \label{ratio}
\end{equation}

This implies that in the strong coupling regime it is more likely
for the system to spontaneously emit out the sides of the cavity
than directly into the cavity mode. Since we only begin a
conditioned field measurement after a cavity mode emission, then
it becomes necessary to consider what happens in the likely event
of a cavity photon following a spontaneous emission.

\begin{figure}
\leavevmode \centering \epsfxsize=3.1in \epsffile{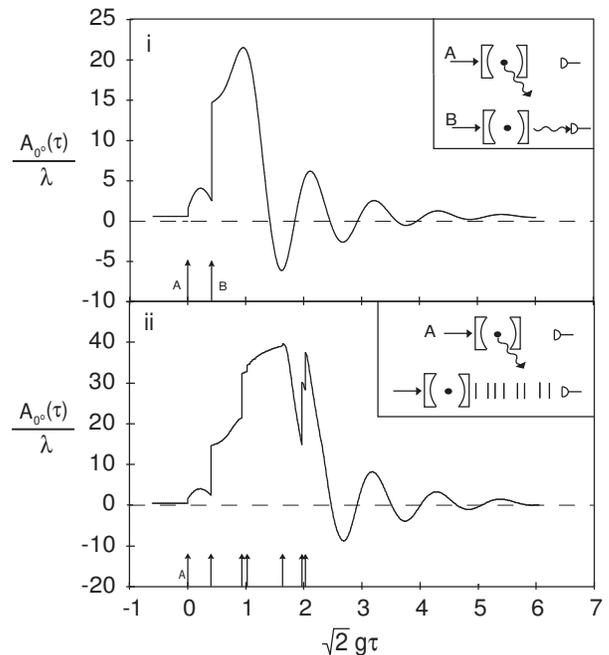}
\caption{A quantum trajectory simulation which shows the time
evolution of the field back to steady state after the detection of
a photon. i. A spontaneous emission event followed by a cavity
emission which starts the averaging process. ii. A spontaneous
emission event followed by many cavity emission events. The inset
shows the sequence of events in terms of the cavity QED system and
the detector. Both figures were prepared with the following
parameters: $N=2$, $X = 18.1$, $(g,\kappa,\gamma)/(2\pi)$ =
$(38.0/\sqrt{2},8.7,3.0)$ Mhz.}\label{fig4}
\end{figure}

Quantum trajectories allow one to follow the dynamics of the
system under higher excitation. Figure \ref{fig4} shows
representative trajectories calculated from Eq.~(\ref{stochschr})
when we have two atoms in the cavity and $r\approx1$. The complete
trajectory is a series of these excursions from the steady state
well separated in time, and a returning to the steady state as
shown. In Fig.~\ref{fig4}i the evolution starts with a spontaneous
emission (A) out the side of the cavity, followed at (B) by a
photon escaping through the cavity mode that gets registered by
the APD detector. Note how the field jumps positive and changes
curvature with the escaping cavity photon. This is exactly the
type of process which gives rise to the incoherent peak in the
spectrum of squeezing.

We offer the following qualitative explanation of these types of
events. The driving field ($\epsilon/\kappa$), atom-field coupling
($g$), and decay rates ($\kappa,\gamma$) are such that the system
is in a regime where the cavity field is bunched. If we have a
spontaneous emission event when the system has few excitations we
return to the steady state value as in Fig.~\ref{fig5}i. If the
spontaneous emission event happens while in the bunched regime,
followed by a cavity emission, then there are probably more
excitations in the system. With one of the atoms removed from the
system following the spontaneous emission, the probability for
this energy to be in the cavity mode is increased. If we detect a
cavity photon soon after the spontaneous emission, then we are
probably in a regime where the intracavity field undergoes a large
amplitude fluctuation, and the value of the cavity field is higher
than the steady state value. This causes an upward jump in the
expectation of the field. This is similar to the upward jump in
the conditioned field of an OPO operated well below threshold
\cite{carmichael93}. In that system the conditioned steady state
is small, since the system is most likely in the ground state,
with a small probability of having a pair of photons in the
cavity. When a photon emerges from the cavity one photon remains,
and the conditioned average field rises. It is also clear that
these types of events increase linearly with the number of atoms
in the cavity, since the ratio of spontaneous emission events to
cavity loss events is $2NC_1$.

If we drive the same system much harder the time evolution of the
conditional field shows multiple jumps, some from spontaneous
emission and some from escapes through the mode. The dynamics get
very complicated and we show an example in Fig.~\ref{fig4}ii for
illustration, but the general trend remains, that the average
value of the field is much larger than the steady state in such
cases.

Figure \ref{fig4} demonstrates the power of studying the single
trajectories and also provides the major result of this paper. We
see that the field will have a noise background arising from the
mean cavity field, on top of other quantum fluctuations, when one
has a spontaneous emission event out the side of the cavity. This
is usually followed by a series of cavity emissions. This effect
depends strongly on the number of atoms in the system and also on
the strength of the driving field. This can not be seen from a
direct numerical solution of the master equation, but now, thanks
to the wave-particle correlator unravelling of the master
equation, we begin to see exactly how these squeezing spectra are
degraded by the incoherent process of spontaneous emission.

\begin{figure}
\leavevmode \centering \epsfxsize=3.1in \epsffile{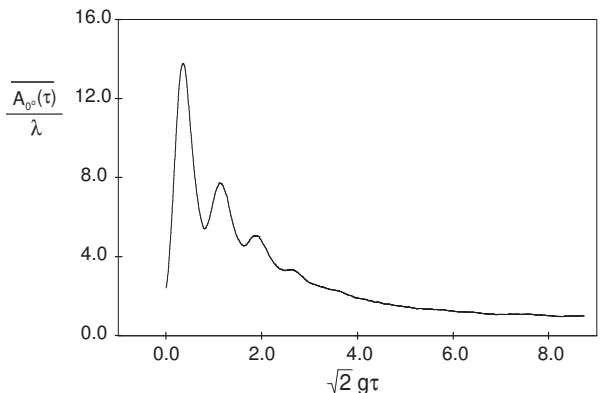}
\caption{Conditioned field evolution calculated from 5000
realizations of a photon counting, quantum trajectory, simulation
for N=2 and high intensity. $X = 18.1$, $(g,\kappa,\gamma)/(2\pi)$
= $(38.0/\sqrt{2},8.7,3.0)$ Mhz.}\label{fig10}
\end{figure}

The entire trajectory is a collection of events well separated in
time of the type in Figs.~\ref{fig4}i and \ref{fig4}ii. If we
average over many random realizations of these different events
with an initial cavity emission setting the trigger at $t=0$, then
we recover the averaged conditioned field evolution. Figure
\ref{fig10} shows such an average field which after symmetrization
and upon taking the Fourier transform leads to a spectrum which
resembles the experimental observations qualitatively. This also
shows a similar behaviour as the oscillation from
Fig.~\ref{fig8}ii and therefore a legitimate connection can still
be made between the correlation function of Eq.~\ref{eqn:prop} and
the spectrum of squeezing slightly beyond the weak field regime.

\subsection{Time Symmetry}

One result that we have only briefly mentioned is the fact that
Fig.~\ref{fig5}i is symmetric with respect to $\tau=0$ while the
single events presented in Figs.~\ref{fig3} and \ref{fig4} show a
clear time asymmetry. This symmetry is recovered as a result of
the BHD back action. The back action comes about because of the
quantum superposition of both the local oscillator and cavity
signal field (see Eq.~(\ref{homm})). It is not correct to think of
these two fields as separable, a classical notion, but that each
time a local oscillator photon is detected, it also effects the
evolution of the cavity field. This implies that the conditional
system state at time $t$ is correlated with the shot noise that
has appeared over the recent past. Thus, through the ``start"
clicks (particle aspect), we postselect a subensemble of the shot
noise which has in effect been filtered through the system's
dynamical response function (Hamiltonian evolution).

Quantum trajectories tend to reveal precisely those physical
attributes they are trying to measure. By adding the homodyne
detector we attempt to measure the fluctuations in the field
amplitude, and when we do this these fluctuations appear in the
averaged photocurrent before the arrival of a triggering photon.

\subsection{Future Work}

By stating the problem in the time domain, we open the
possibilities to apply quantum feedback theory
\cite{weisman93,doherty00} in order to conditionally stabilize the
atom-cavity system evolution against decoherence from the
environment. This stabilization will not be optimal since the
information obtained is limited to the quasimode leaking out of
the cavity and any spontaneous emission event will be missed in
the feedback loop. This calls for operating the cavity QED system
in the low intensity regime to minimize spontaneous emission. We
are currently investigating the theoretical and experimental
requirements to develop such a program.

\section{Conclusions}

The conditional evolution of the field
of a cavity QED system in the strong coupling regime shows
non-classical behavior. This is purely quantal in nature and
reflects the intrinsic dynamics triggered by the collapse of
the wavefunction after the detection (conditioning) of a photon.

Looking at the time evolution of the conditional state as it
regresses back to steady state we find that its behavior is far
from random. When uninterrupted by other events, it follows a
dynamic path that shows the coupling between the atom and the
cavity and the significant excursion away from equilibrium brought
back by a single quantum measurement. The spectrum of squeezing
registers exactly the same behavior, but without the time and
phase information critical to understanding the dynamics of the
measurement process.

As we implement the measuring device used in the laboratory, we
require an averaging process to observe the conditional
evolution of the electromagnetic field. That evolution is
non-classical. The phase is exactly $\pi$ from the driving
field and it ocurrs when the intensity has a fluctuation up. This
is contrary to all classical intuition that would say that then
the field should have also a maximum, but in fact the field
shows a minimum.

The presence of the homodyne measurement introduces
back action into the system that then provides the foundation
for the time symmetric wave-particle correlation function
when third order correlations can be neglected.

The presence of other avenues for the interruption
of the evolution back to steady state, such as spontaneous emission,
modify the correlation function, decreasing its non-classicality, while
at the same time modify the spectrum of squeezing around
the origin with a peak that shows just the empty cavity width for
the one atom case, or the spontantous emission width once the
system is allowed to have more than one atom.

The wave-particle correlation function opens the door to further
the study of quantum optical systems in the time domain that are
necessary to understand better the possible realizations of
quantum feedback systems.

\section{Acknowledgements}

Work supported by NSF.

\end{document}